\documentclass[preprint]{aastex}

\newcommand{\Msun}{ $M_{\odot}$}
\newcommand{\ebv}{$E$$($$B$$-$$V$$)$}
\newcommand{\Fabs}{$F_{abs}$}
\newcommand{\Fem}{$F_{em}$}
\newcommand{\peryr}{yr$^{-1}$}
\newcommand{\tb}{($T_d$, $\beta$)}
\newcommand{\fWm}{$\cdot$10$^{-15}$Wm$^{-2}$}
\newcommand{\Ha}{H$\alpha$}
\newcommand{\kmsMpc}{km~s$^{-1}$Mpc$^{-1}$}

\slugcomment{To appear in the Astronomical Journal, May 2001}

\shorttitle{Energy Budget of MS 1512-cB58}
\shortauthors{Sawicki}

\begin{document}

%% LaTeX will automatically break titles if they run longer than
%% one line. However, you may use \\ to force a line break if
%% you desire.

\title{The Ultraviolet-Far Infrared Energy Budget of the Gravitationally
Lensed Lyman Break Galaxy MS~1512--cB58}

\author{Marcin Sawicki} 
\affil{California Institute of Technology, Mail Code 320-47, Pasadena,
California 91125, USA}
\email{sawicki@mop.caltech.edu}

\begin{abstract}

A 2-hour service-mode SCUBA observation of the gravitationally-lensed
Lyman break galaxy MS~1512--cB58 resulted in a 3$\sigma$ upper limit
of 3.9 mJy at 850\micron.  A comparison of this upper limit with
values expected from rest-UV/optical measurements of extiction
suggests that dust temperature ($T_d$) and/or emissivity index
($\beta$) in cB58 may be substantially higher than is seen in local
galaxies, or that the attenuation curve in cB58 may be even gentler
than the already quite mild SMC dust law.  If dust temperature $T_d$
and emissivity index $\beta$ in cB58 are similar to those seen in
local IRAS-seleceted galaxies, then cB58's dust mass is
$M_d$$\lesssim$$10^{7.7}$\Msun\ and its star formation rate is
SFR$\lesssim$10\Msun\peryr\ (for $q_0$=0.1, $H_0$=75 \kmsMpc).  This
SFR upper limit is lower than the star formation rate measured from
\Ha, thus giving further support to the notion that \tb\ values in
cB58 are higher than those seen in local galaxies.  It thus appears
that our understanding of dust in this extensively studied Lyman break
galaxy is poor, and observations at other wavelentghs are needed to
better understand dust at high redshift.  Such observations can be
provided by the upcoming SIRTF mission for which cB58's expected flux
densities are calculated.
\end{abstract}


\section{INTRODUCTION}

\subsection{Dust in star-forming galaxies at $z \approx 3$}\label{intro_general}

The recent discovery of a large population of $z \approx 3$
star-forming Lyman break galaxies (LBGs; e.g., Steidel et al.\ 1996,
1999) has generated enormous enthusiasm in the field of galaxy
formation and evolution.  Among other developments, luminosities of
$z$$\approx$3 galaxies have been used to trace out the history of
cosmic star formation (e.g., Lilly et al.\ 1996) to a time when the
Universe was only $\sim$ 10 \% of its present age (e.g., Madau et al.\
1996, 1998; Sawicki, Lin, \& Yee, 1997; Steidel 1999).  At face value,
these studies suggest that star formation in the universe rises with
lookback time until flattening or even decreasing sometime after $z
\approx 1$.  However, this picture of cosmic star formation is
complicated by the possibility that substantial amounts of dust may
be present in LBGs.  Specifically, star formation rates (SFRs) at high
redshift are usually inferred from rest-UV luminosities (typically
rest-1500\AA), but UV light is strongly absorbed by
dust. Consequently, if significant dust is present, UV luminosities
will be significantly underestimated, resulting in potentially
dramatic underestimates of star formation rates.

The amount of dust in LBGs and its effect on the inferred star
formation rates remains uncertain.  Estimates of absorption in LBGs
range between factors of 3 and 20 at rest-1500\AA, with a
corresponding range of correction factors applied to estimates of star
formation rates (e.g., Meurer et al.\ 1997, 1999; Pettini et al.\
1998; Sawicki \& Yee 1998).  The large range in the estimates of dust
attenuation is a result of the different assumptions made regarding
both the shape of the underlying spectral energy distributions (SEDs)
and, especially, of the dust attenuation laws in Lyman break galaxies.
Given the range of possible attenuation corrections that results from
reasonable assumptions, it is unlikely that a consensus about the
amount of stellar light intercepted by dust will be reached on the
basis of rest-frame UV and optical data alone.

A complementary approach to the rest-frame UV-optical studies is that
of measuring not the amount of energy absorbed by dust, but the amount
re-emitted by it.  Because of conservation of energy, these two
quantities should be equal, and so observing thermal dust emission in
the rest-frame far infrared (FIR) should help constrain the amount of
energy absorbed by dust in the rest-frame UV.  Thus by observing
thermal dust emission one can infer the amount of UV flux absorbed by
dust and hence the amount of star formation in high-$z$ galaxies.

The commissioning of the sensitive Submillimetre Common-User
Bolometer Array (SCUBA; Holland et al.\ 1999) on the James Clerk
Maxwell Telescope (JCMT) has resulted in a number of what are
effectively deep blank-field surveys (e.g., Smail, Ivison, \& Blain,
1997; Barger, Cowie, \& Sanders 1999; Lilly et al.\ 1999) as well as
two studies which specifically targeted Lyman break galaxies (Ouchi et
al.\ 1999; Chapman et al.\ 2000).  Ouchi et al.\ (1999) considered the
fact that none of the 17 spectroscopically-confirmed LBGs in the
northern Hubble Deep Field (HDF) have been detected in an ultra-deep
SCUBA image of the HDF \cite{hug98}.  Using the SCUBA non-detections
together with fits to optical and near-infrared (NIR) photometry, they
concluded that LBGs are likely similar to low-reddening starbursts in
the local Universe, but that their dust temperature is $T_d \gtrsim
40$K (for a dust emissivity index $\beta$=1) --- higher than typically
found in local galaxies.  Chapman et al.\ (2000) observed a targeted
sample of 16 LBGs, but detected only one object.  They concluded that
predictions from UV colors underestimate LBG 850\micron\ flux
densities by a factor of two or more, except in the case of their sole
detection which is associated with an abnormally red and hence,
presumably, abnormally dusty outlier.  The non-detections of both
these studies suggest that typical Lyman break galaxies are beyond the
reach of the best sub-mm instrument currently available.

\subsection{The gravitationally-lensed Lyman break galaxy MS 1512--cB58}\label{intro_cB58}

The $z$=2.72 Lyman break galaxy MS~1512--cB58 (cB58 hereafter) was
discovered serendipitously behind a rich $z$=$0.37$ cluster of
galaxies (Yee et al.\ 1996).  Gravitational lensing by the cluster
potential and by the cluster cD galaxy located only 6\arcsec\ away,
results in an amplification of cB58's light by a factor of 22--40
(Seitz et al.\ 1998), thus making it possible to study cB58 in detail
that is impossible for other Lyman break galaxies.

A number of detailed studies of cB58 have been undertaken to date, of
which three have addressed the issue of dust: Ellingson et al.\ (1996)
compared rest-UV and -optical photometry of cB58 with spectral
synthesis models and concluded that cB58 is dominated by a young
stellar population whose light suffers from \ebv\ $\approx$ 0.3 of
extinction (a factor of $\sim$11 at rest-1500~\AA) under the
assumption of a LMC-like extinction law.  Teplitz et al.\ (2000)
compared the observed ratios of Balmer lines with those expected on
theoretical grounds, and concluded that attenuation due to dust is
\ebv\ = 0.27 for a LMC-like attenuation law (a factor of 9 at
1500~\AA).  Finally, Pettini et al.\ (2000) used a very deep
rest-frame UV spectrum of cB58 in order to study in detail its stellar
population; on the basis of the slope of cB58's UV continuum and the
assumption of a relatively young stellar population, they concluded
that for an LMC-like extinction law cB58 suffers from a factor of 7 of
absorption at rest-1500~\AA\ (\ebv\ = 0.24).

While for the same attenuation law the three different estimates of
cB58's extinction agree with each other to better than a factor of
two, the picture is complicated by the issue of cB58's unknown
attenuation curve. It is not known what absorption law is appropriate
for Lyman break galaxies, yet different local absorbtion laws can give
vastly different results.  For example, in the Pettini et al.\ (2000)
study, the LMC-like absortion curve results in a factor of 7
absortption at 1500~\AA, an SMC-like curve gives only a factor of 3,
while the Calzetti (1997) curve (which is appropriate for local
starburst galaxies) gives a factor of 20.  Thus, results from the same
measurement span a range of a factor of six in attenuation depending
on what attenuation curve is assumed.  Until the shape of the high
redshift attenuation law is better understood, UV-based estimates of
extinction in Lyman break galaxies will have to remain uncertain at
this level.

As was mentioned in \S\ref{intro_general}, complementary insights into
the nature of dust at high redshift can be gleaned from observations
of the dust's thermal emission.  Gravitational magnification allows
cB58 to be observed at rest-frame FIR to a sensitivity unattainable
for normal, unlensed, Lyman break galaxies, and so a relatively short
(2-hour integration) service-mode observation of cB58 was carried out
with SCUBA on the JCMT.  This observation, which resulted in a
sensitive upper limit on the 850\micron\ flux density, is briefly
described in \S\ref{observations}.  In \S\ref{budget} the SCUBA
observation is combined with data from the optical and near-IR to
produce an overall energy budget for cB58 and to help constrain the
nature of its dust.  The star formation rate in cB58 and the mass of
its dust are constrained in \S\ref{sfrdust}, while the detectability
of cB58 with the Space InfraRed Telescope Facility (SIRTF) is
discussed in \S\ref{prospects}.  The main results are summarized in
\S\ref{conclusions}.
 
\section{SCUBA DATA ON cB58}\label{observations}

\subsection{Data}

The galaxy cB58 was observed in service mode with Sub-millimetre
Common-User Bolometer Array on the JCMT and a total integration of
7200 seconds was obtained in SCUBA's photometric mode on 1998 June 26
and 1998 July 2.  During observations the instrument was nodded in
order to subtract sky and flux calibration was done on both dates by
observing Uranus.  Atmospheric optical depth values were similar on
both dates and the two sets of data were combined after scaling by the
gains.  The data were reduced by JCMT staff using standard techniques.

No 850\micron\ signal was detected at cB58's location above the
3$\sigma$ sky noise of 3.9 mJy.  This upper limit is of similar
sensitivity as, and is in agreement with the 4.2$\pm$0.9 mJy value
reported recently in a conference paper by van der Werf et al.\
(2000).  Note that because of the presence of a foreground cD galaxy
within the SCUBA beam {\it any} 850\micron\ SCUBA observation of cB58
has to be regarded as an upper limit (see \S~\ref{caveats}).  Thus,
throughout the rest of this paper, 3.9~mJy is adopted as the upper
limit on cB58's 850\micron\ flux density.

\subsection{Caveats}\label{caveats}

While the gravitational amplification provided by the foreground
cluster and its cD galaxy make it possible to reach deep sensitivity
limits in moderate amounts of observing time, they also introduce two
complications which need to be noted.

The first complication is that for further analysis it is necessary to
assume that the gravitational amplification is the same for the
optical/NIR and for sub-mm light.  This assumption is reasonable given
that the young stars which are responsible for the rest-UV light are
likely well mixed on galactic scales with the dust which is
responsible for absorbing their UV light and re-emitting it in the
FIR.  However, were this not the case, then the assumption that the
sub-mm flux is amplified by the same factor as the rest-UV and
-optical light would not be true and would complicate the comparison
of optical and sub-mm fluxes presented in \S\ref{budget}; it
would also put in doubt the calculations of FIR limits on the star
formation rate and dust mass in cB58 (\S\ref{sfrdust}), as these
calculations assume that the gravitational amplification of cB58's
sub-mm light can be estimated from the optical.

The second complication is introduced by the presence 6\arcsec\ away
of the cD galaxy in the centre of the cluster MS1512.  This foreground
object is within the 15\arcsec\ diffraction-limited beam of the JCMT
at 850\micron\ and --- given that cD galaxies at $z \sim 0.4$ can
produce appreciable sub-mm emission (e.g., Edge et al.\ 1999) --- may
be contributing to the 850\micron\ flux density.  Therefore it should
be kept in mind that {\it any} SCUBA observation of cB58 --- even one
which reports a detection, such as that by van der Werf et al.\ (2000)
--- has to be regarded as an upper limit on that object's flux
density.  Throughout the analysis that follows, it will be assumed
that the upper limit on cB58's 850\micron\ flux density is 3.9 mJy.  This
assumption is conservative in that it is consistent with both the
non-detection reported above and with the van der Werf et al.\ (2000)
value, and does take into account any possible contributions from the
neighbouring cD galaxy.

\section{THE UV-FIR ENERGY BUDGET OF cB-58}\label{budget}

In this Section, the 3.9 mJy upper limit on cB58's 850\micron\ flux
density will be compared with expectations based on observations of
the rest-UV and -optical regions of its SED.  The comparison will be
done in terms of observed quantities --- fluxes and flux densities,
rather than luminosities and luminosity densities --- thereby avoiding
the need to assume a specific cosmology and lensing magnification.  In
\S\ref{Fabs}, recent UV-based measurements of extinction in cB58 will
be used to estimate the amount of rest-UV flux that is absorbed by
dust.  Then (in \S\ref{Fem}) the expected 850\micron\ flux density
will be calculated using the assumption that all the flux absorbed at
rest-UV is re-radiated in the rest-FIR as a modified blackbody.  In
\S\ref{local_comp} this expeced 850\micron\ flux density will be
compared with the 3.9 mJy SCUBA upper limit to constrain the
properties of dust in cB58.

\subsection{Flux absorbed by dust in the rest-UV}\label{Fabs}

\begin{figure}
\plotone{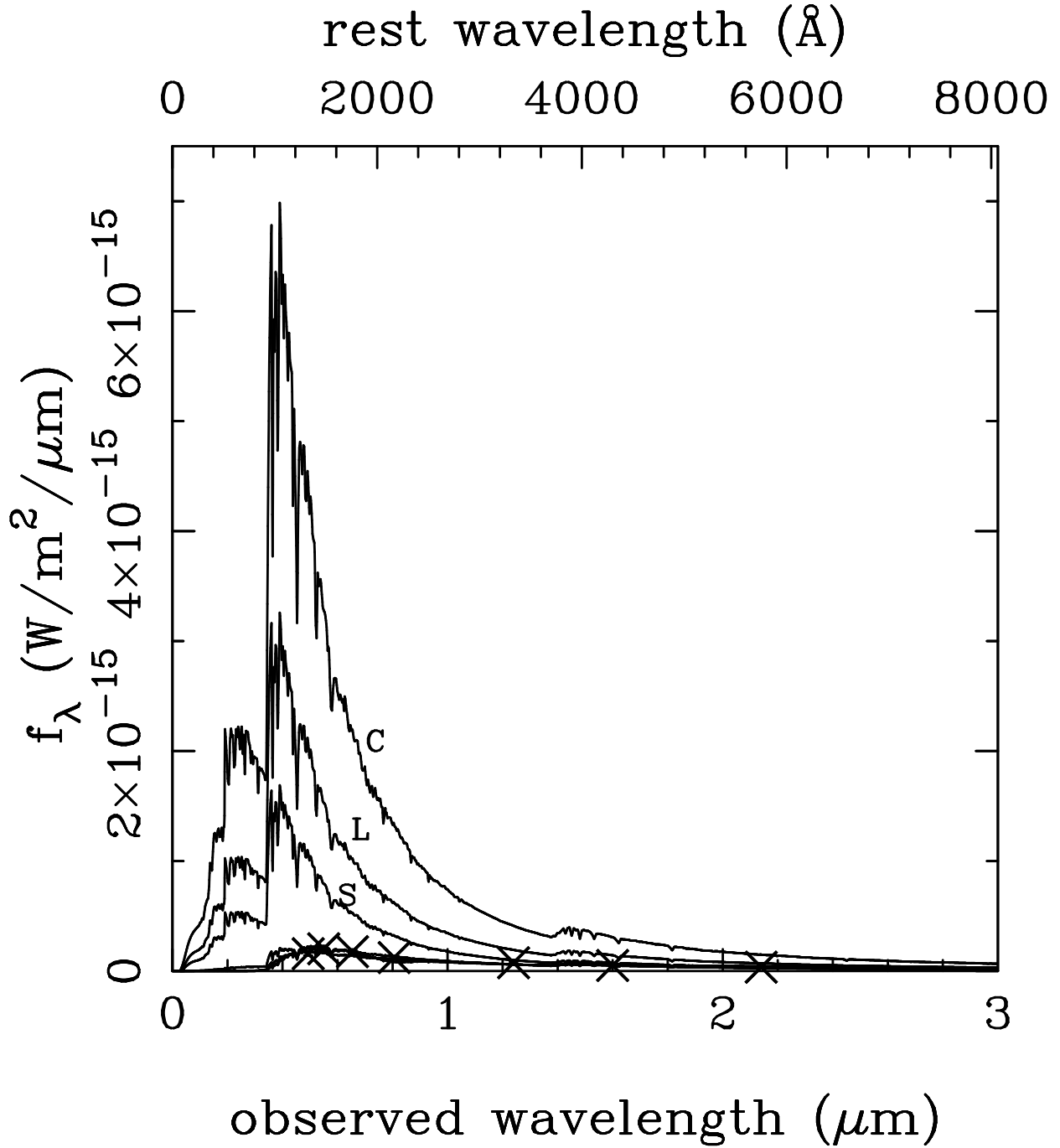}
\caption{\label{optical_SEDs.fig} Attenuated and unattenuated SED
curves for cB58.  Crosses show the optical and IR photometry of cB58
from Yee et al.\ (1996) and Ellingson et al.\ (1996).  The curves that
pass through the crosses are Bruzual and Charlot (1993) models
adjusted for dust attenuation using the Calzetti (1997), LMC, and SMC
extinction laws with extinction values from Pettini et al. (2000),
while the three upper curves, labelled ``C'', ``L'', and ``S'', are
the same Bruzual \& Charlot models but uncorrected for extinction.
Areas between the upper, unattenuated, curves, and the corresponding
lower, attenuated ones, are the amounts of flux absorbed by dust (see
Eq.~\ref{Fabs.eq}).}
\end{figure}

To estimate cB58's expected sub-mm flux density, $S_{850\mu m}$, it is
first necessary to calculate $F_{abs}$, the amount of flux that is
absorbed at rest-frame UV and optical wavelengths.  The calculation of
\Fabs\ was done by comparing the dust-free and dust-attenuated model
SEDs of young stellar populations (1996 version of the Bruzual \&
Charlot 1993 models) after normalizing them to cB58's broadband
photometry (Yee et al., 1996; Ellingson et al., 1996).  Symbolically
this calculation can be expressed as
\begin{equation}\label{Fabs.eq} 
F_{abs} = n \int^{\infty}_{0} [f_\lambda(\lambda) - a(\lambda)
f_\lambda(\lambda)] d\lambda,
\end{equation}
where $f_{\lambda}$ is the unattenuated SED, $a(\lambda)$ is the
extinction as a function of wavelength for the assumed \ebv\ value and
extinction curve, and $n$ is the normalization obtained by scaling the
attenuated SED to the broadband photometry (see also
Fig.~\ref{optical_SEDs.fig}).

The specific values of the various parameters, such as the amount of
extinction, age of the stellar population, metallicity, and stellar
initial mass function, were as follows (see also
Table~\ref{Fabs.tab}).  Three extinction laws were considered --- the
SMC law of Bouchet et al.\ (1985), the Fitzpatrick (1986) LMC law, and
the Calzetti (1997) law appropriate for local starburst galaxies.
\ebv\ values were taken from Pettini el al.\ (2000), and are \ebv=0.1
for the SMC law, \ebv=0.24 for the LMC law, and \ebv=0.29 for the
Calzetti law; the Pettini et al.\ (2000) values were adopted as of the
three extinction measurements for cB58 (Ellingson et al. 1996; Pettini
et al.\ 2000; Teplitz et al.\ 2000) they give the lowest extinction
corrections and are thus the most conservative (\S~\ref{intro_cB58}).
The underlying SEDs were assumed to be dominated by young stellar
populations since the UV-optical light in cB58 is dominated by young
stars (Ellingson et al.\ 1996, Pettini et al.\ 2000); specifically,
models with ages of 10, 100, and 255 Myr were used.  Both
instantaneous burst models and constant SFR models were initially
considered, but it was found that for the 100 and 255 Myr
instantaneous burst models the fit between the model SED and the
broadband photometry was extremely poor and so for the instantaneous
burst case only the 10 Myr model was retained.  Different
metallicities and initial mass functions (IMFs) were considered at
first, but it was found that \Fabs\ does not depend strongly on either
metallicity or IMF; for definitiveness, metallicity was taken to be
0.4$Z_{\odot}$ and the Salpeter (1955) IMF was adopted.

As can be seen in Table~\ref{Fabs.tab}, the values of \Fabs\ for the
Calzetti attenuation law cluster around \Fabs\ = 2.5\fWm, those for
the LMC law are around 1.0\fWm, and those for the SMC law are
typically around or slightly above 0.5\fWm.  Motivated by these
typical \Fabs\ values, let us define three cases for dust attenuation,
namely ``case C'', corresponding \Fabs=2.5\fWm, ``case L'',
corresponding \Fabs=1.0\fWm, ``case S'', corresponding
\Fabs=0.5\fWm\ (see bottom of Table~\ref{Fabs.tab}).  These three
cases correspond broadly to the effects of the three different
attenuation laws.

\subsection{Expected 850\micron\ flux density}\label{Fem}

With the estimate of \Fabs\ in hand, the expected flux density at
850\micron\ can be calculated by assuming conservation of energy
(\Fabs=\Fem) and a dust emissivity curve.  Specifically, let's assume
that the dust radiates as an optically thin modified blackbody,
$\{1-exp[-(\nu/\nu_0)^\beta]\} B_\nu (T_d)$, where $B_\nu$ is the
Planck function and $\nu_0 = c/\lambda_0$ is the critical frequency at
which dust becomes optically thin.  The emitted flux is then the
integral of the redshifted emissivity curve,
\begin{equation}
F_{em}=n'{\int_{0}^{\infty} \{1-\exp[-(\nu/\nu_0)^\beta]\}
B_{\nu/(1+z)} (T_d) d\nu },
\end{equation}
where $n'$ is the constant of proportionality which incorporates
cosmological terms and which is evaluated by requiring that \Fem=\Fabs.
The expected 850\micron\ flux density is then
\begin{equation}\label{Snu.eq}
S_{\nu} (850\mu m) = n'\{1-\exp[-(\nu(850 \mu
m)/\nu_0)^\beta]\} B^{850\mu m}_{\nu/(1+z)}(T_d),
\end{equation}
where the redshifted modified blackbody is evaluated at the observed
wavelength of 850\micron.

\begin{figure}
\plotone{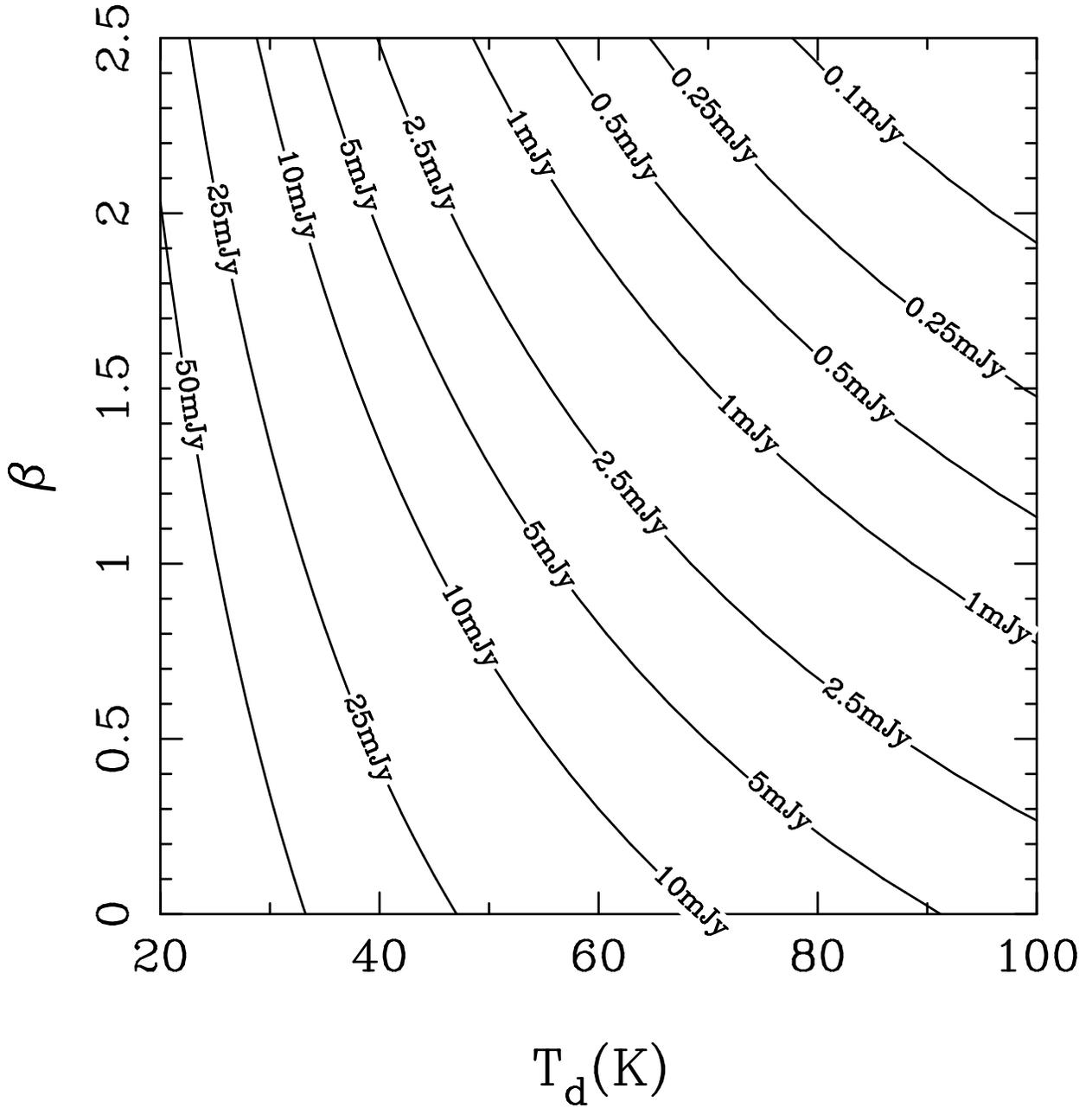}
\caption{\label{case2_850um.fig} Expected 850\micron\ flux density for
the illustrative case of SMC-like attenuation (``case S'') that is
optically thin everywhere in the FIR.  Note that since the
normalization of Eq.~(\ref{Snu.eq}) depends on the rest-UV absorption
only through the value of $F_{abs}$, the $S_{\nu (850\mu m)}$ values
for cases L and C can easily be read off from the Figure by
multiplying the plotted flux density values by 2 (for case L) or 5
(for case C).  }
\end{figure}

Figure~\ref{case2_850um.fig} shows the expected values of 850\micron\
flux density for case S dust attenuation (i.e., for \Fabs = 0.5\fWm)
and under the assumption that dust is optically thin everywhere
($\nu_0\rightarrow\infty$, or $\lambda_0\rightarrow 0$).  Note that
for any given $S_\nu(850\mu m)$ value, dust temperature and emissivity
index are degenerate, so it would be impossible to separate $T_d$ from
$\beta$ even if a SCUBA 850\micron\ detection were available.  For a
SCUBA {\it upper limit}, it is only possible to rule out {\it areas}
of \tb\ parameter space: for a given upper limit value, the area to the
right of the corresponding curve is permitted whereas that to its left
is ruled out. 

\subsection{Comparison of cB58 to galaxies in the local Universe}\label{local_comp}

\begin{figure}
\plotone{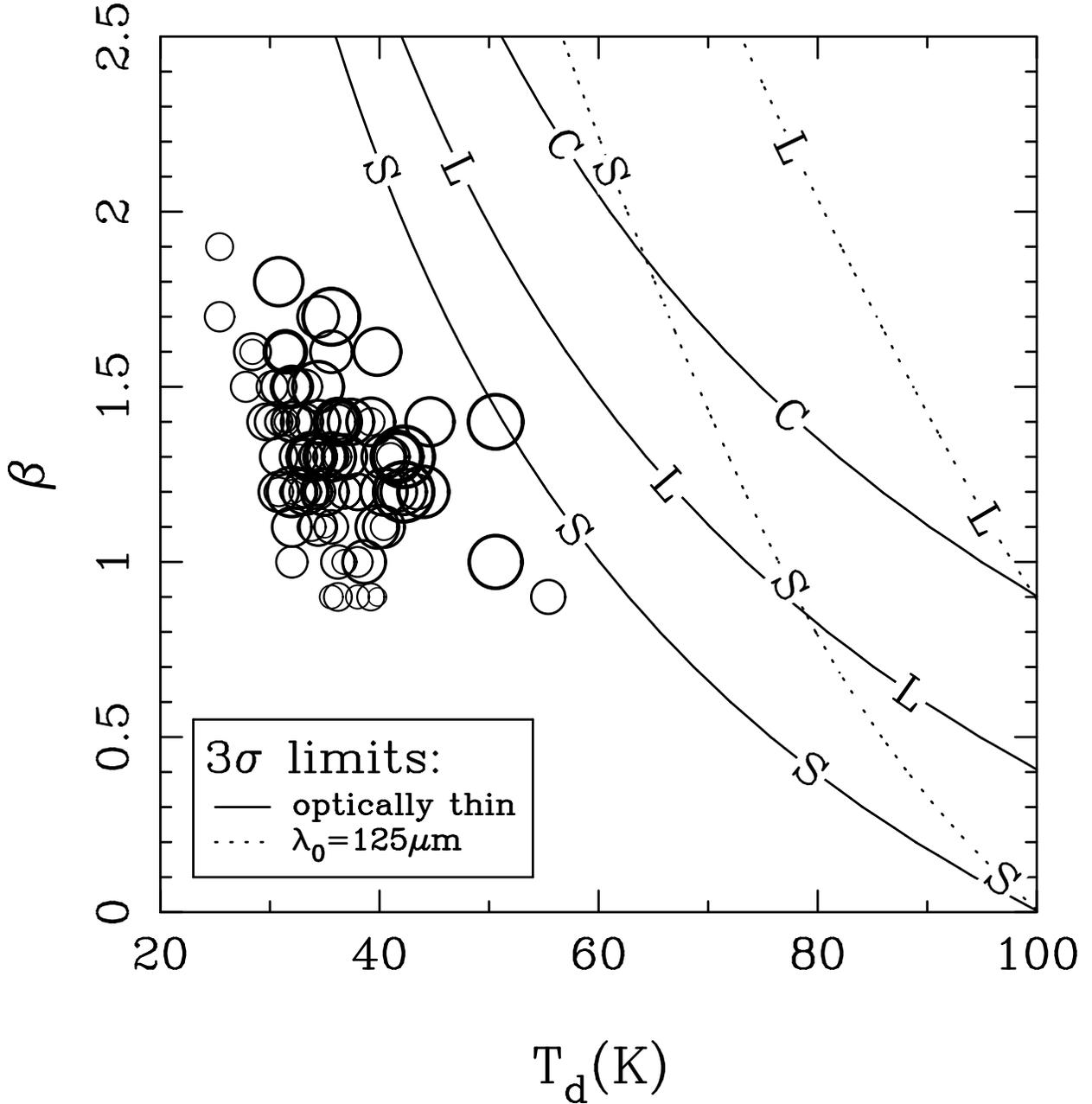}
\caption{\label{f850_Tbeta_limits.fig} Regions in $T_d-\beta$ space
allowed by the combination of SCUBA 850\micron\ upper limit and
rest-frame UV/optical data.  The solid and broken lines show the
3$\sigma$ upper limits, with the allowable regions being to the upper
right of each curve.  The solid curves are for dust that is optically
thin at all wavelengths, while the broken curves are for dust that
becomes optically thick at $\lambda_0$=125\micron.  The curves
labelled ``S'' are for the case of SMC-like extinction in the UV (see
Table~\ref{Fabs.tab} for the definition), those labelled ``L'' are for
LMC-like dust, and those labelled ``C'' are for Calzetti-like dust.
Open circles represent local galaxies from the survey of Dunne et al.\
(2000), with symbol size proportional to FIR luminosity.  Note that
the local galaxies lie outside the regions permitted for cB58,
suggesting that dust properties in cB58 are unlike those in the local
sample.  }
\end{figure}

Figure~\ref{f850_Tbeta_limits.fig} shows the constraints in the \tb\
plane imposed by the SCUBA 3.9 mJy upper limit for different
assumptions about dust absorption and emission laws.  Specifically,
areas to the upper right of the {\it solid curves} are the values of
$T_d$ and $\beta$ allowed in the case of dust emission that is
optically thin everywhere ($\lambda_0\rightarrow 0$).  Labels on the
curves indicate case L, S, and C dust, while open circles show values
of \tb\ measured by Dunne et al.\ (2000) for 104 local galaxies from
the IRAS Bright Galaxy Sample (Soifer et al.\ 1989).  The restrictions
on the \tb\ parameter space become more severe when $\lambda_0$$>$0,
as is illustrated for $\lambda_0$=125\micron ({\it broken lines}).

No galaxy from the local, IRAS-selected sample of Dunne et al.\ (2000)
can match the \tb\ values required of cB58 for any of the dust
absorption cases (C, L, S) and $\lambda_0$ values.  The discrepancy is
least severe when dust emission is assumed to be optically thin
everywhere ($\lambda_0\rightarrow 0$) and becomes worse for the more
realistic case of $\lambda_0 > 0$. Furthermore, the discrepancy is
strongest for case C dust, which corresponds to the Calzetti
extinction curve which is derived for starburst galaxies and could
thus be thought to be the most appropriate extinction curve for cB58.
Thus, for all the absorption laws considered here, the optical-NIR
data require so much flux to be absorbed in the rest-frame UV that it
cannot be re-emitted in the rest-FIR at dust temperatures and $\beta$
values typical of galaxies in the local universe (see also van der
Werf et al.\ 2000).

A number of possibilities exists which may account for this apparent
disagreement between the \tb\ values required of cB58 and those seen
in local galaxies:

\begin{enumerate}

\item It is possible that dust in cB58 is indeed hotter than dust seen
in local IRAS selected galaxies.  A similar conclusion was reached for
Hubble Deep Field LBGs by Ouchi et al.\ (1999), who concluded that
their dust temperatures are $T_d$$\gtrsim$40K under the less robust
assumption of a Calzetti attenuation curve.  The discrepancy in
dust temperature between cB58 and local galaxies may be due to
evolution in dust properties between $z$$\approx$3 and today.
Alternatively, it could be that cB58 is a very $L_{FIR}$-luminous
member of a continuum of galaxies whose less luminous members have
more normal dust temperatures.

\item An alternative is that a significant component of {\it hot} dust
is present in cB58.  No attempt has been made here to fit a
two-temperature dust model because to do so data at multiple sub-mm
wavalengths would be needed.  However, if significant amount of the
overall FIR flux is radiated via a hot dust component then the
850\micron\ flux density would be lower than predicted with the simple
single-temperature models of \S\ref{Fem}.  Indeed, Ouchi et al.\
(1999) estimate that in some local galaxies as much as a third of the
overall FIR flux may be radiated by hot dust.  However, the
discrepancy between the SCUBA limit for cB58 and local galaxies is
generally larger than a factor of 2 --- only in the most favourable
case S optically thin dust could the 850\micron\ SCUBA limit be
reconciled with {\it some} members of the local galaxy population and
even then only with one or two outliers in the 104 galaxies in the
Dunne et al.\ (2000) sample.  It is thus unlikely that the presence of
a hot dust component can account for the entire discrepancy, unless
substantially more flux is radiated by hot dust than by cold dust, in
which case we revert to the the conclusion in item (1) above.

\item A third possibility is that $\beta$ is substantially higher than
in local galaxies.  It is possible to have $T_d$ similar to those
observed in local galaxies (see Fig.~\ref{f850_Tbeta_limits.fig})
provided $\beta$$\gtrsim$2.5.  While such high $\beta$ values are not
seen in local galaxies, it is possible that dust at high redshift has
unusual composition which might result in such a high emissivity
index.

\item Another possibility is that the attenuation laws (SMC, LMC,
Calzetti) used in \S\ref{Fabs} to compute \Fabs\ are inappropriate for
cB58.  If the attenuation curve in cB58 were more mild than even
the SMC law, then the expected flux densities would be overestimated.
A milder extinction curve could restore agreement between \tb\ in
cB58 and in the local sample.  However, if such a mild attenuation
curve is required in Lyman break galaxies, then even the lowest,
factor-of-3, estimates of the effect of dust on star formation rates
in LBGs and on the cosmic star formation rate density at high redshift
would have to be revised {\it downwards}.

\item The final possibility considered is that at least some of the
sub-mm flux comes from a different region of cB58 than the optical
light.  If this were the case then one might expect the sub-mm and
optical fluxes to be uncorrelated, especially so if lensing amplified
the two components by different factors.  However, given that there is
already a {\it deficit} of rest-FIR light, any {\it additional}
FIR-emitting component would only make the discrepancy worse.
Alternatively, it is also extremely improbable that the dust which
absorbs the rest-UV radiation is lensed more strongly than the dust
which re-emits it.  Thus the deficit seen in the sub-mm cannot be
explained as due to emission from a component which is different from
that responsible for absorption in the UV.

\end{enumerate}

It thus appears that dust in cB58 cannot be easily modeled by a simple
model of attenuation and single-temperature emission unless one
assumes \tb\ values vastly different from those seen in local
galaxies, or a very mild dust attenuation law.  Data at other FIR
wavelengths are needed to better constrain the shape of the dust
emission curve and hence the properties of dust in cB58.
Possibilities for obtaining such data in the near future are briefly
explored in \S\ref{prospects}.  However, first the analysis of the
data already in hand will be completed by calculating the SCUBA limits
on cB58's star formation rate and mass of dust.

\section{LIMITS ON STAR FORMATION RATE AND DUST MASS}\label{sfrdust}

Assuming that the sub-mm continuum in cB58 is due to thermal emission
from single-temperature dust heated by massive young stars, the 3.9
mJy upper limit on the 850\micron\ flux density can be used to
constrain both the star formation rate and the mass of dust in
that galaxy.

\subsection{Star formation rate}\label{sfr}

The star formation rate of cB58 can be constrained using the limit on
the 850\micron\ flux density.  Following Hughes et al.\ (1997):
\begin{equation}
SFR = \frac{1}{f_{lens}} \epsilon 10^{-10} \frac{L_{FIR}}{L_{\sun}} M_{\sun} yr^{-1},
\end{equation}
where it is assumed that the lensing magnification of cB58's flux is
$f_{lens} = 30$ (Seitz et al.\ 1998 give the allowable range of
22--40).  Hughes et al.\ 1997 give $\epsilon$=0.8--2.1, and here the
value of $\epsilon$=2.1 is adopted because using this high value will
give a conservative, robust upper limit on the star formation
rate.  The far-IR luminosity, $L_{FIR}$, is calculated by integrating
the modified blackbody in the observed frame,
\begin{equation}\label{LFIR.eq}
L_{FIR} = 4 \pi D_L^2 \int_{(1+z) 10\mu m}^{(1+z)2mm} [\nu/(1+z)]^\beta B_{\nu/(1+z)} (T_d) d\nu,
\end{equation}
and normalizing using the constraint that the 850\micron\ flux density
is $<$3.9 mJy.  Here, $D_L$ is the usual luminosity distance, given by
\begin{equation}
D_L = \frac{2c}{H_0 \Omega_0^2} [ \Omega_0 z + (\Omega_0 - z) (\sqrt{\Omega_0 z + 1} -1)].
\end{equation}
Note that the calculation of $L_{FIR}$ (Eq. \ref{LFIR.eq}) depends on
the dust temperature and emissivity index $\beta$ and that, therefore,
the constraint on the SFR will depend on the assumed values of these
(unknown) parameters.  Note also that because Eq.\ (\ref{LFIR.eq}) is
normalized using an upper limit on the 850\micron\ flux density, the
calculated SFR will also be an upper limit.

\begin{figure}
\plotone{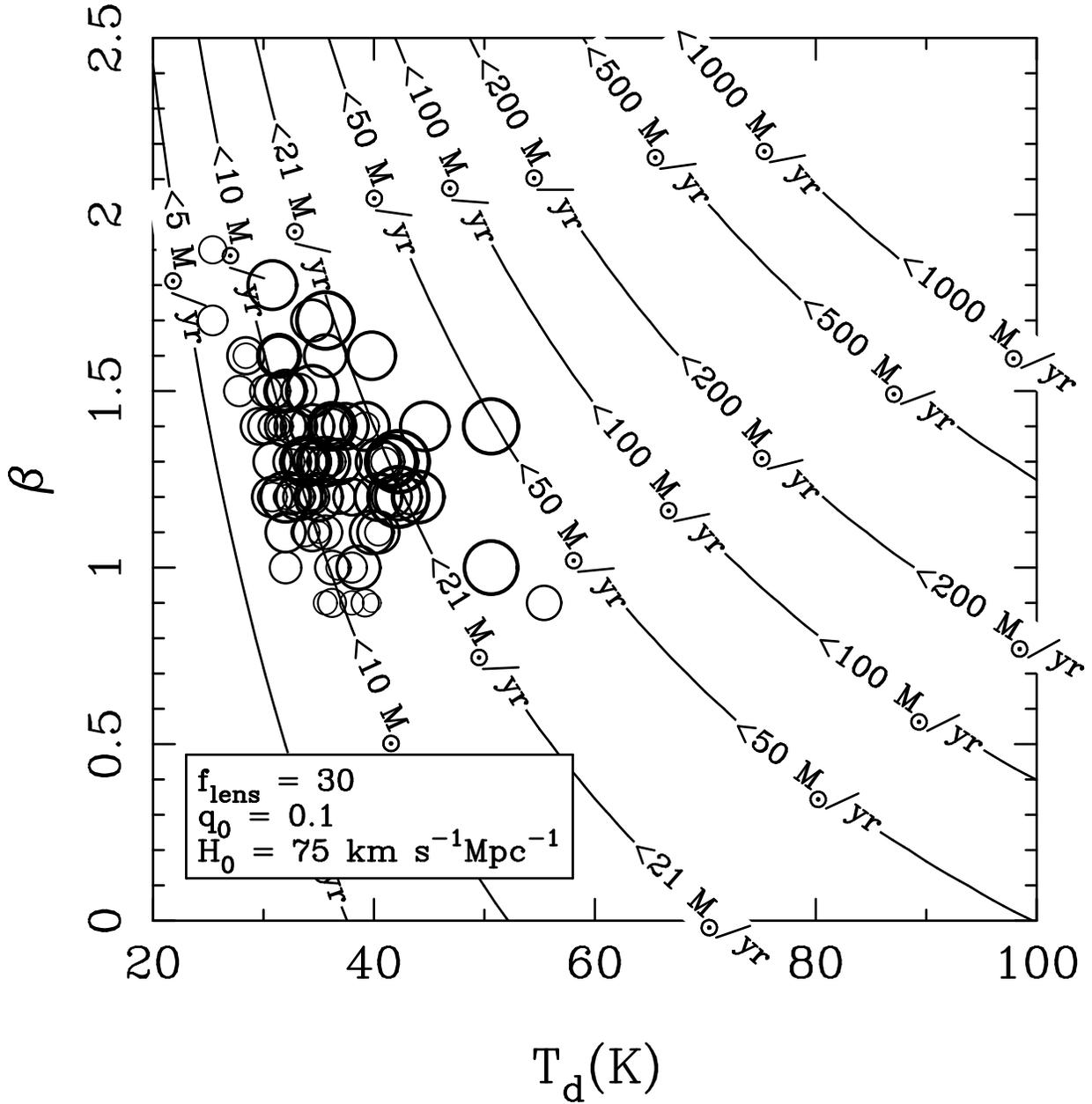}
\caption{\label{SFR.fig}Sub-mm limits on the star formation rate in
cB58. 3$\sigma$ upper limits for the adopted cosmology and lensing
magnification are shown as a function of dust temperature $T_d$ and
emissivity index $\beta$.  To be consistent with the Teplitz et al.\
(2000) SFR$_{H\alpha}$, cB58's \tb\ values must be to the upper right
of the $<$21\Msun/yr line, and are thus consistent with only a few of
the 104 local galaxies in the sample of Dunne et al.\ (2000).}
\end{figure}

Figure \ref{SFR.fig} shows the constraints on cB58's star formation
rate as a function of dust temperature $T_d$ and emissivity index
$\beta$ and for an assumed cosmology of $q_0=0.1$ and $H_0$ = 75
\kmsMpc.  If cB58 were comparable to a typical local IRAS-selected
galaxy then it would have $T_d$=36K and $\beta$= 1.3 (mean values for
the sample of Dunne et al., 2000) and would thus have a star formation
rate of $\lesssim$ 10 \Msun\peryr.  This upper limit on the star
formation rate is lower than the (lensing-corrected) value of 21
\Msun\peryr\ derived by Teplitz et al.\ (2000) from cB58's \Ha\ flux.
Indeed, to be consistent with the Teplitz et al.\ SFR$_{H\alpha}$
value, cB58 must be located to the upper right of the $<21$\Msun/yr
line in Fig.~\ref{SFR.fig} --- i.e.  at \tb\ values higher than those
seen in typical local galaxies.  Note that this argument, which is
based on comparing the \Ha-based SFR with the SCUBA upper limit on the
850\micron\ flux density, is independent of the argument based on the
comparison of rest-UV extinction models with the SCUBA limit
(\S\ref{budget}) and yet results in the same conclusion that the \tb\
values in cB58 are high and unlike those seen in local galaxies.  It
thus strengthens the conclusion of \S\ref{local_comp} that dust in
cB58 appears to by unlike that in local IRAS-selected galaxies.

\subsection{Mass of dust}\label{Mdust}

As with the star formation rate, the mass of dust can be constrained
using the limit on the 850\micron\ flux density.  Following Hughes et
al.\ (1997), the mass of dust can be expressed as
\begin{equation}
M_d = \frac{1}{f_{lens}}\frac{S^{obs} D_L^2} {(1+z) k_d^{rest}
B(\nu ^{rest}, T_d)},
\end{equation}
where $S^{obs}$ is the constraint on the 850\micron\ flux density
(i.e., $S^{obs} < 3.9$ mJy), $z=2.72$ is the redshift of cB58, and
$B(\nu ^{rest},T_d)$ is the value of the Planck function evaluated at
the rest frequency which corresponds to the observed wavelength of
850\micron.  The mass absorption coefficient $k_d^{rest}$ is quite
uncertain, with different estimates of $k_d^{rest}$(800\micron)
ranging from 0.04 m$^2$kg$^{-1}$ to 0.3 m$^2$kg$^{-1}$ (Hughes et al.\
1997); here, $k_d^{rest}$(800\micron)=0.04 is adopted, since this low
$k_d^{rest}$ gives a conservative, robust upper limit on $M_d$.  To
extrapolate $k_d^{rest}$(800\micron) to the observed rest-wavelength
of 850$/(1+z)$\micron, it is assumed that
$k_d^{rest}$$\propto$$\lambda^{-\beta}$.

\begin{figure}
\plotone{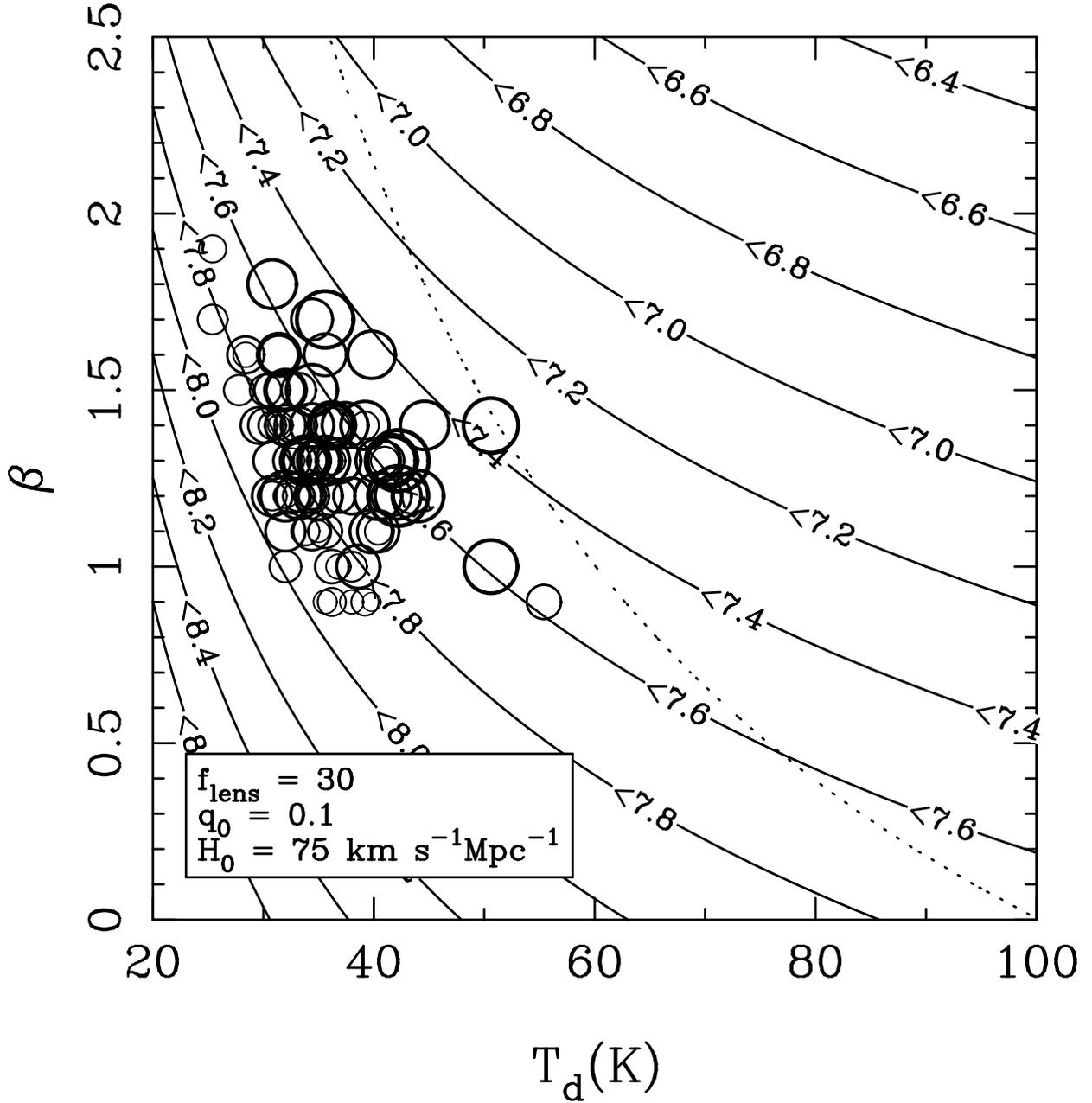}
\caption{\label{Mdust.fig}Limits on the mass of dust.  The solid
lines, labelled in units of log($M_d/M_\sun$), indicate the 3$\sigma$
limits on the mass of dust for the adopted cosmology and lensing
magnification.  The area of \tb\ space to the right of the broken line
is allowed by the (least restrictive) case S optically thin dust.  }
\end{figure}

Figure~\ref{Mdust.fig} shows the limits on cB58's dust mass for a
range of values of temperature and $\beta$ and for a $q_0$=0.1,
$H_0$=75 \kmsMpc\ cosmology.  Assuming that \tb\ values in cB58 are
similar to the mean of the local sample of Dunne et al.\ (2000), the
mass of dust in cB58 is $M_d$$\lesssim$$10^{7.7}$\Msun.  If, on the
other hand, it is assumed that \tb\ values in cB58 are higher than is
seen in local IRAS-selected galaxies, then the upper limit on the dust
mass drops somewhat to, for example, $M_d$$\lesssim$$10^{7.3}$\Msun,
for \tb=(60K, 1.3).

\section{PROSPECTS FOR THE (NEARBY) FUTURE}\label{prospects}

\begin{figure}
\plotone{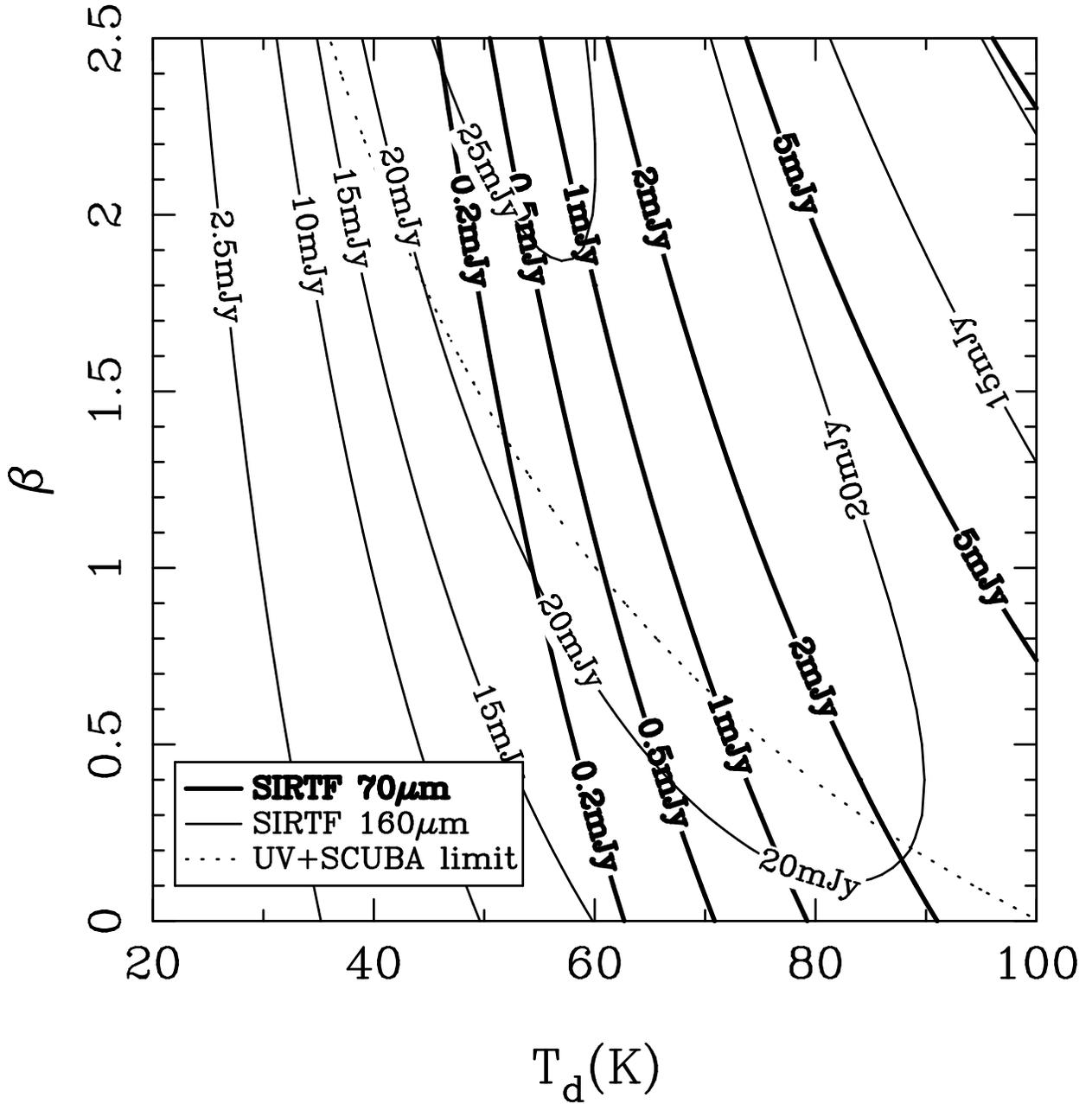}
\caption{\label{SIRTF}Expected flux densities in the SIRTF 70\micron\
and 160\micron\ bands, for case S, optically thin dust.  The area to
the upper right of the dotted line is the region allowable by the
UV-FIR constraints in the case of the (least restrictive) optically
thin case S dust. Flux density values for cases L and C dust can be
easily read off from this figure by multiplying the flux density
values plotted by 2 and 5, respectively.}
\end{figure}

As was discussed in \S\ref{local_comp} and \S\ref{sfr}, dust
properties in cB58 appear to be different from those seen in local
IRAS-selected galaxies.  To constrain the shape of cB58's dust
emissivity curve, observations at additional rest-frame FIR
wavelengths are needed.  Such observations will be possible with the
Space Infrared Telescope Facility (SIRTF) which is scheduled for
launch in mid-2002.  Figure~\ref{SIRTF} shows the flux densities
expected in the 70\micron\ and 160\micron\ SIRTF bandpasses from case
S optically thin dust.  The area to the upper right of the dotted
curve is the region allowable by the comparison of optical-NIR and
SCUBA data.  Thus, the thermal dust contribution to the 70\micron\
flux density is likely to be $\gtrsim$0.5 mJy, which is attainable at
5$\sigma$ in $\sim$ 4 hours of integration\footnote{SIRTF
sensitivities are taken from \tt
http://sirtf.caltech.edu/SciUser/MIPS/html/mipspht\_sens.html}.
Unless $T_d$ is above 100K, cB58 can be expected to have a flux
density of $\gtrsim$10 mJy at 160\micron, which will be detectable to
very high significance in only a few minutes of SIRTF time.
Therefore, cB58 should be easily detectable with SIRTF at both
70\micron\ and 160\micron.  Such multiwavelength FIR observations will
help constain the shape of the emissivity curve in cB58, and will thus
yield important insights into the properties of dust in this
prototypical Lyman break galaxy and, by extension, into the true
amount of star formation at $z$$\approx$3.

\section{CONCLUSIONS}\label{conclusions}

A 2-hour SCUBA observation of the Lyman break galaxy MS 1512--cB58
yielded a 3$\sigma$ upper limit of 3.9 mJy at 850\micron.  Given that
cB58 is gravitationally lensed by a factor of $\sim$30, it is thus not
surprising that virtually no other Lyman break galaxies have been
detected in the sub-millimeter even in very deep surveys (Hughes
et al.\ 1998; Chapman et al.\ 2000).

The SCUBA upper limit on cB58's 850\micron\ flux density is
surprisingly low when compared with values expected on the basis of
rest-frame UV/optical observations (\S\ref{budget}).  This discrepancy
suggests that values of dust temperature and/or emissivity index
$\beta$ may be substantially higher in cB58 than in local
IRAS-selected galaxies.  Alternatively, the attenuation curve in cB58
may be gentler even than the quite mild SMC dust law, resulting in
substantially less absorption in the rest-UV (factor of $\lesssim 3$
at 1500\AA) than is commonly assumed (Pettini et al.\ 2000; Teplitz et
al.\ 2000).  In either case, it is clear that the understanding of
dust in this extensively studied Lyman break galaxy is poor.

The SCUBA upper limit constrains the mass of dust in cB58 to
$M_d$$\lesssim$$10^{7.7}$\Msun\ for a $q_0$=0.1, $H_0$=75 km/s/Mpc
cosmology and a lensing magnification of $f_{lens}$=30
(\S\ref{Mdust}).  If \tb\ values similar to those seen in local
galaxies are adopted for cB58, then its star formation rate is
SFR$\lesssim$10\Msun\peryr (\S\ref{sfr}).  This upper limit on cB58's
star formation rate is lower than the star formation rate measured
from \Ha\ by Teplitz et al.\ (2000) and thus gives further support to
the notion that \tb\ values in cB58 are higher than those in local
galaxies.

This analysis illustrates that properties of dust in Lyman break
galaxies are not well understood.  Further multiwavelength studies
will be needed to understand the shape of the dust emissivity curve in
cB58 and in other LBGs.  Estimating the flux emitted by dust in cB58
from that absorbed in the rest-frame UV, it is found that cB58 should
be easily detectable at 70\micron\ with SIRTF, and --- with a few
hours' integration --- at 160\micron.  Using such multiwavelength
studies to compare the picture of dust in absorption with that of dust
in emission may eventually lead to an understanding of dust in Lyman
break galaxies that is lacking at present.

%%%

I thank Tracy Clarke, Gabriela Mall\'en-Ornelas, Gerry Neugebauer, Tom
Soifer, and Nick Scoville for discussions and comments.  Special
thanks are due Henry Matthews of the Joint Astronomy Centre who
carried out the SCUBA observations and reductions for this service
observing project. The James Clerk Maxwell Telescope is operated by
the Joint Astronomy Center on behalf of the UK Particle Physics and
Astronomy Research Council, the Netherlands Organization for
Scientific Research, and the National Research Council of Canada.
Financial support for this work came from the Natural Sciences and
Engineering Research Council (NSERC) of Canada and from NSF grant
AST-9618686.

%% Generally speaking, only the figure captions, and not the figures
%% themselves, are included in electronic manuscript submissions.
%% Use \figcaption to format your figure captions. They should begin on a
%% new page.

\clearpage

%% No more than seven \figcaption commands are allowed per page,
%% so if you have more than seven captions, insert a \clearpage
%% after every seventh one.

%% There must be a \figcaption command for each legend. Key the text of the
%% legend and the optional \label in curly braces. If you wish, you may
%% include the name of the corresponding figure file in square brackets.
%% The label is for identification purposes only. It will not insert the
%% figures themselves into the document.
%% If you want to include your art in the paper, use \plotone.
%% Refer to the on-line documentation for details.

%% Tables should be submitted one per page, so put a \clearpage before
%% each one.

%% Two options are available to the author for producing tables:  the
%% deluxetable environment provided by the AASTeX package or the LaTeX
%% table environment.  Use of deluxetable is preferred.
%%

\clearpage

\begin{deluxetable}{lcccrc}
\tablewidth{0pt} \tablecaption{\label{Fabs.tab}Flux absorbed by dust
for different SED models and dust laws} 
\tablehead{ \colhead{Dust} & \colhead{history\tablenotemark{a}} & \colhead{age\tablenotemark{b}} & \colhead{\Fabs}\\
            \colhead{}     & \colhead{} &                         \colhead{(Myr)}                & \colhead{($10^{-15}$Wm$^{-2}$)}
}
\startdata
SMC, \ebv=0.10 		& ssp 	& 10 	& 0.5	\\
			& cons	& 10	& 0.8 	\\
		 	& cons	& 100	& 0.6	\\
			& cons	& 255	& 0.5	\\
LMC, \ebv=0.24		& ssp 	& 10 	& 0.9	\\
			& cons	& 10	& 1.3	\\
		 	& cons	& 100	& 1.0	\\
			& cons	& 255	& 0.9	\\
Calzetti, \ebv=0.29 	& ssp 	& 10 	& 2.1	\\
			& cons	& 10	& 3.2	\\
		 	& cons	& 100	& 2.4	\\
			& cons	& 255	& 2.0	\\
\tableline		
Case S			& 	& 	& 0.5   \\
Case L			& 	& 	& 1.0   \\
Case C			& 	& 	& 2.5   \\
\enddata
\tablenotetext{a}{History of star formation. Following Bruzual \& Charlot (1993), cons refers to a constant star formation 
history and ssp refers to an instantaneous burst of star formation}
\tablenotetext{b}{Age since the onset of the episode of star formation 
that dominates the optical-IR SED.}
\end{deluxetable}

%% The following command ends your manuscript. LaTeX will ignore any text
%% that appears after it.

\end{document}